\documentstyle[12pt]{article}
\begin{document}

\title{Black-Scholes equation from Gauge Theory of Arbitrage}

\author{
Kirill Ilinski$^{1,2}$
\thanks{E-mail: kni@th.ph.bham.ac.uk}
\ and \ Gleb Kalinin$^{1}$
\thanks{E-mail: kalinin@snoopy.phys.spbu.ru}
\\ [1cm]
{\small\it $^1$ IPhys Group, CAPE,
14-th line of Vasilievskii's Island, 29}\\
{\small\it St-Petersburg, 199178, Russian Federation}
\\ [0.2cm] \\ [0.3cm]
{\small\it $^2$ School of Physics and Space Research,
University of Birmingham,} \\
{\small\it Edgbaston B15 2TT, Birmingham, United Kingdom}
}

\vspace{3cm}
\date{ }
\vspace{1cm}
\maketitle

\begin{abstract}
We apply Gauge Theory of Arbitrage (GTA)~\cite{hep-th/9710148} to
derivative pricing. We show how the standard results of Black-Scholes
analysis appear from GTA and derive correction to the Black-Scholes
equation due to a virtual arbitrage and speculators' reaction on it. The
model accounts for both violation of the no-arbitrage constraint and
non-Brownian price walks which resemble real financial data. The
correction is nonlocal and transforms the differential Black-Scholes
equation to an integro-differential one.
\end{abstract}

\section{Introduction}
The Black-Scholes equation~\cite{BS} for prices of derivative financial
instruments is probably the most beautiful rigorous result in the
theory of financial analysis. It surprises not only with a simple
derivation but also with its high applicability which might be even more
important.

The derivation is based on several natural simplifications (see Appendix
1 for an one-page introduction to the financial derivatives). Among those
the geometrical Brownian motion (or, more general, quasi-Brownian motion)
model for the price of an underlying asset and the no-arbitrage
constraint are most difficult to avoid. Below we stop to describe each of
the assumptions in more detail.

Quasi-Brownian walks as a model for price increments keep many convenient
features, such as easy-to-fit parameters~\cite{Dubofsky}, a developed and
handleable mathematical description~\cite{Duffie} which results in the
existence of exact solutions obeying explicit
equations~\cite{BS,Wilmott}, and indeed resemble real price movements.
However, the resemblance is not perfect and many deviations have been
observed~\cite{fin,Mantegna,BS94,TLF}. These studies reveal other
characteristic processes which substitute for Brownian random walks.
These are the Truncated Levy Flights which are regularized Stable Levy
processes slowly converging to the Brownian process at large times and
large prices~\cite{SM}. Another, more popular, approach to model price
movements with excess kurtosis is to use the quasi-Brownian walks but
with a stochastic variance (volatility)~\cite{Hobson}. A recent study
\cite{Liu} shows that, at least for the S\&P500 index, the volatility is
distributed log-normal and follows the geometrical random walk with a
good measure of accuracy. However, the last approach cannot explain
scaling properties and correlations observed in financial data.

A number of approaches have been developed to correct the Black-Scholes
analysis for the case of deviations from a simple geometrical Brownian
motion model for prices. The well-known and widely accepted way to do it
is to generalize the Black-Scholes analysis to quasi-Brownian motions
with the stochastic volatility. Other approaches use various versions of
risk-neutral treatment which differ by the details of the averaging
procedure. A recent paper~\cite{TLF} presents an attempt to account for
the Truncated Levy Flight nature of underlying asset price walks based on
a Bouchaud and Sornette scheme~\cite{BS94,risk}.

The present paper suggests the different approach to the problem which
releases both no-arbitrage assumption and Brownian character of the price
walks. As was shown in Ref~\cite{ISfin1} based on the Gauge Theory of
Arbitrage (GTA) picture, the active trade behavior of speculators who profit
from temporary mispricing of assets may be one of the reasons for the
deviation of the probability distribution function (PDF) from the
log-normal law and a possible explanation of memory effects. The origin
of this is the fact that for a mispriced asset there are cash flows
(inflow for a undervalued asset and outflow for an overvalued asset)
which shifts the price due to a kind of supply-demand mechanism. This
shift is {\it directed} and results in damping of the mispricing which, in
its own, reduces the width of the PDF peak and increases the peak height.
On the other hand, the arbitrageurs create an additive noise which causes
the power-law wings of the PDF~\cite{Takayasu,Sornette}. The
investigation of the influence of the directed speculations on the
distribution function of prices in a very simple gauge arbitrage model
demonstrated a qualitative agreement with the observed
data~\cite{ISfin1}. Thus, it was demonstrated that the model distribution
function of the price increments is similar to one observed in a real market
and possesses the same scaling properties. This motivates us to
apply developed formalism to derivative pricing and examine corrections
to Black-Scholes equation due to arbitrageurs. Being responsible for both
the deviation of the probability distribution function of the underlying
asset prices and, at the same time, restoring virtually violated
no-arbitrage constraints, speculators must cause corrections to the
Brownian walks based arbitrage-free equation for prices of derivatives.
The goal of the paper is to derive these corrections in a simple gauge
arbitrage model which generalizes the GTA stock exchange model.

We will not go into details of GTA here and redirect the reader to
Refs~\cite{hep-th/9710148,ISfin1}. However, to make the exposition
self-contained, the main issues of GTA are sketched below.

\subsection{GTA in brief}
Let us remind the reader of concepts and notations of
GTA~\cite{hep-th/9710148} used in our consideration. This is
field-theoretical description of the virtual arbitrage possibilities and
corresponding money flows. In this framework the net present value (NPV)
calculation and asset exchanges are interpreted in geometrical terms
as parallel transports in some sophisticated fibre bundle space. This
allows us to map the capital market theory to a theory similar to
electrodynamics and then use the machinery of quantum field theory. It
was shown that the free quantum gauge theory is equivalent to the
assumption of log-normal walks for assets prices. In general, the theory
 resembles electrodynamics where particles with positive charge
("securities") and negative charges ("debts") interact with each other
through the electromagnetic field (arbitrage excess return). In the case
of a local virtual arbitrage opportunity money flows into the profitable
security. Entering positive charges and leaving negative ones screen the
profitable fluctuation and restore the equilibrium, i.e. speculators wash
out the arbitrage opportunity. The starting point for GTA construction is
an understanding of Net Present Value calculations as a parallel
transport in some fibre bundle. The NPV states that money has a time
value. This time value has to be taken into account through so-called
discounting procedure. If an amount of money $F$ is to be received in $T$
years' time, the Present Value of that amount ($NPV(F)$) is the sum of
money $P$ (principal) which, if invested today, would generate the
compound amount $F$ in $T$ years' time (for simplicity $r$ is considered
constant over the $T$ years):
$$
NPV(F) \equiv P = \frac{F}{(1+r)^T} \ .
$$

The interest rate involved in this calculation is known as the discount
rate and the term $(1+r)^{-T}$ is known as T-year discount factor $D_T$:
\begin{equation}
D_T = (1+r)^{-T} \ .
\label{DT}
\end{equation}
Thus NPV method shows how to compare money amounts came at different
moments of time~\cite{Lumby}.

This last phrase points directly the geometrical interpretation
which we use: discounting procedure plays the role of a "parallel
transport" of an amount of money  through time (though in fixed currency).
The discounting factor (\ref{DT}) is then an element of a structural
group of a fibre bundle and the discount rate coincides with the time
component of the connection vector field. The "space" components of the
connection are related to exchange rates and prices. Indeed, exchange
rates and prices are responsible for converting money in different
currencies or different securities, i.e. points of discrete "space", to
the same currency (point of the space) at a fixed moment of time.
They can be interpreted as elements of the structural group which
"transport" the money in "space" directions and are space analogues of
the discount factor. Summing up, the capital market theory has a
geometrical structure which allows us to map it onto a theory of a fibre
bundle.

The curvature tensor of the connection field is related to the arbitrage
which is an operational opportunity to make a risk-free
profit~\cite{Wilmott} with a rate of return higher than the risk-free
interest rate accrued on deposit. As was derived in
Refs~\cite{hep-th/9710148}, the rate of excess return on an elementary
arbitrage operation (a difference between rate of return on the
operation and the risk-free interest rate) is an element of the curvature
tensor calculated from the connection. It can be understood keeping in
mind that the curvature tensor element is related to a difference between two
results of infinitesimal parallel transports performed in different order
with the same initial and final points or, in other words, a gain from an
arbitrage operation. Due to this geometrical interpretation it is
possible to say that the rate of excess return on an elementary
arbitrage operation is an analogue of the electromagnetic field.

In the absence of any uncertainty and money flows, the only state that is
realized is the state of zero arbitrage. However, if we introduce the
uncertainty to the game, prices and rates move and some virtual
nonequivalent possibilities to get more than less appear. Therefore we
can say that the uncertainty play the same role in the developing theory
as the quantization did for quantum gauge theory.

Money flow fields appear in the theory as "matter" fields which are
transported by the connection (interests and exchange rates). It means
that the matter fields interact through the connection. Dilatations of
money units (which do not change real wealth) play the role of gauge
transformation which eliminates the effect of the dilatation by a
corresponding gauge transformations of the connection in the same way as
the Fisher formula does for the real interest rate in the case of an
inflation~\cite{Lumby}. {\it The symmetry of the real wealth to a local
dilatation of money units, security splits and the like is the gauge
symmetry of the theory}.

An investor's strategy is not always optimal. This is due to partially
incomplete information available, partially because of an investor's
internal objectives~\cite{Lumby}. It means that the money flows are not
certain and fluctuate in the same manner as prices and rates do. So
this requires a statistical description of money flows which, once
again, returns us to an effective quantization of the theory.

At this stage we would like to clarify the following misunderstanding
which might emerge. The arbitrage itself implies a possibility to perform
an operation with a risk-free rate of return which is higher than, say, a
bank deposit interest rate. In this sense buying shares cannot be
considered as a such operation because of assumed random walk of the
share price and the corresponding risk. What do we mean then talking
about the arbitrage?  The randomness of the price is equivalent to a
quantization as we explained in~\cite{hep-th/9710148} and the rate of
return on an (arbitrage) plaquette operation is now a quantum variable
which cannot be taken as a complex number. This exactly resembles a
situation with an electromagnetic field which, after the quantization, is
not a number but a quantum variable. However, it does not stop us using
the same name for the variable, imagining virtual quantum fluctuations
and describing the influence of these fluctuations on electric charges,
keeping in mind the calculation of corresponding matrix elements. In the
same way we understand the arbitrage rate of return in the financial
setting.

Summing up, it was shown how to map the capital market to a system of
particles with positive, "securities", and negative, "debts", charges which
interact with each other through an electromagnetic field, the gauge
field of the arbitrage. In the case of a local virtual arbitrage
opportunity, cash flows into the region of configuration space (money go
in the profitable security) while "debts" try to escape from the region.
This brings in positive charges and pushes out negative ones, leading to an
effective screening of the profitable fluctuation. These processes
restore an equilibrium and erase the arbitrage opportunity.

Formalization of the scheme drawn above leads to the lattice
quantum field theory~\cite{Creutz}. At this point the standard machinery of
quantum field theory can be applied to obtain various observables such as
distribution functions of the interest/exchange rates,
response functions of the system and others. It may answer questions about
dynamical response of a financial market, the dynamical portfolio theory
and other problems.

In conclusion we want to add that notions of the (stochastic)
differential geometry appeared in the context of financial modelling in
paper~\cite{Hughston}. It contains several ideas which are similar to
GTA.
\vspace{1.5cm}

In this paper we want to apply the idea to derivative pricing. More
precisely, we show how the standard results of Black-Scholes analysis
appeared from GTA and derive correction to the Black-Scholes equation due
to a virtual arbitrage and speculators reaction on it. This will model
both violation of the no-arbitrage constraint and non-Brownian price
walks which resemble recently observed data.

The paper is organized as follows. In next section we formulate a GTA
model for a description of derivative instruments. To do this we
construct base space of the theory which, in contrast to GTA model for the
simplest stock exchange~\cite{ISfin1}, contains a double ladder. This
corresponds to simultaneously treating cash, shares and derivatives. 
Section 3 is devoted to an investigation of the classical
limit of the action which leads to the Black-Scholes equation in the
particular case of free gauge field dynamics. We show that the plaquette
diagrams obtained in Section 2 have a very simple interpretation as an
imbalance between the derivative and the Black-Scholes hedging portfolio.
Section 4 is devoted to description of the money flow fields and the 
corresponding correction to an effective
action for the derivative price coming from virtual speculations. In the
classical limit it gives a correction to the Black-Scholes equation. 
The last section completes the paper with final remarks.
In the appendix we give simple derivation of the Black-Scholes equation.

\section{GTA model of derivatives}
In this section we construct a GTA model for a share-cash-derivative
system. To simplify the consideration we consider only one type of
shares and the perfect capital market conditions are implied.
The consideration of this paper is not restricted by any type of
concrete derivative contracts. However, to simplify the consideration
we will illustrate the GTA application by an analysis of European and
American call options.

Shares or derivatives can be exchanged with cash and vice versa. The
corresponding exchange rates are $S_i$ and $C_i$ (one share or derivative
contract is exchanged on $S_i$ or $C_i$ units of cash) at some moment
$t_i$, and the reverse rates (cash to share or derivative) are $S^{-1}_i$
and $C^{-1}_i$. We consider period from starting point $t=0$ up to moment
$t=T$. For the case of option we assume that $T$ is expiration time. We
suppose that there exists a shortest interval of time $\Delta=T/N$ and this
$\Delta$ is taken as a unit time. So, the exchange rates $S_i$ and $C_i$ are
quoted on a set of equidistant times: $\{t_i\}_{i=0}^N, t_i =i\Delta$ and
represent the parallel transport along legs of a double ladder base
graph. The interest rate for cash is $r_b$ so that between two subsequent
times $t_i$ and $t_{i+1}$ the volume of cash is increased by factor
$e^{r_b \Delta}$. The shares and derivatives are characterized by rates
$r_1$ and $r_2$ correspondingly. These rates realize parallel transport
in time direction.

Following Refs~\cite{hep-th/9710148} we consider elementary arbitrage
operations when an arbitrageur borrows one share at time $t_i$,
sells it for $S_i$ units of cash, put the cash in the bank until time
$t_{i+1}$ and, at time ($t_{i+1}$), closes his short position
borrowing $e^{r_1\Delta}S_{i+1}$ units of cash and buying shares. The
result of the operation for the arbitrageur will be
$e^{r_b\Delta}S_{i}-e^{r_1\Delta}S_{i+1}$ units of cash. The excess
return on this operation is
\begin{equation}
Q^{(1)}_i = S_i e^{r_b\Delta}S^{-1}_{i+1} e^{-r_1\Delta} - 1\ .
\label{npv1}
\end{equation}
To get this expression we discounted the amount and converted it in shares
since the operation was started in the shares. Equation (\ref{npv1})
has a form of the curvature tensor element corresponding to drawing
assets through the cycle. If $Q^{(1)}_i\ne0$ an arbitrageur can get excess
return performing this or the reverse operation. The following quantity
\begin{equation}
R^{(1)}_i = (S^{-1}_i e^{r_1\Delta}S_{i+1} e^{-r_b\Delta}
+ S_i e^{r_b\Delta}S^{-1}_{i+1} e^{-r_1\Delta} - 2)/\Delta
\label{eq:1}
\end{equation}
is used to measure the arbitrage (excess rate of return) on local
cash-share operation. The absences of the arbitrage is equivalent to the
equality
$$
S^{-1}_i e^{r_1\Delta}S_{i+1} e^{-r_b\Delta} =
S_i e^{r_b\Delta}S^{-1}_{i+1} e^{-r_1\Delta} = 1\ .
$$
The same can be done for another possible arbitrage operation. For
cash-derivative plaquette it gives the following quantities
\begin{equation}
Q^{(2)}_i = C_i e^{r_b\Delta}C^{-1}_{i+1} e^{-r_2\Delta} - 1\ ,
\label{npv2}
\end{equation}
\begin{equation}
R^{(2)}_i = (C^{-1}_i e^{r_2\Delta}C_{i+1} e^{-r_b\Delta}
 + C_i e^{r_b\Delta}C^{-1}_{i+1} e^{-r_2\Delta}
 - 2)/\Delta \ .
\label{eq:2}
\end{equation}

Looking at (\ref{eq:1},\ref{eq:2}) we can conclude that the arbitrage
is represented in the theory by the curvature of the connection.
Precisely, in the continuous limit ($\Delta\to0$) the RHS of
Eqns~(\ref{eq:1},\ref{eq:2}) converges as usual to a square of the
curvature tensor element multiplied by area of plaquette
\begin{equation}
R^{(l)}_i = (Q^{(l)}_i)^2/\Delta\ .
\label{pq}
\end{equation}
Curvature is $Q^{(l)}_i/\Delta$, the area of plaquette is proportional to
the shortest time interval $\Delta$ and fixed "space" length which is omitted.

Being represented by the curvature tensor, the notion of the arbitrage as well as
the quantity $R^{(l)}_i$ are gauge invariant. All of this allows us to say
that the rate of excess return on an elementary arbitrage operation is
an analogue of the electromagnetic field. In the absence of uncertainty
(or, in other words, in the absence of random walks of prices, exchange and
interest rates) the only state realized is the state of zero
arbitrage. However, if we introduce the uncertainty, prices and the
rates move and some virtual arbitrage possibilities appear. Therefore,
we can say that uncertainty plays the same role in the developing
theory as the quantization does for the quantum gauge theory.

Further development of the consideration are based on the following
assumptions about the dynamics~\cite{hep-th/9710148}: gauge
invariance, locality, correspondence principle, extremal action
principle, limited rationality (uncertainty) and absence of correlation
between excess returns on different plaquettes. Formally, these
assumptions are summed up in the functional form of the probability
$P(\{S_i,r_k\})$ to find a set of the exchange rates/interest rates
$\{S_i,r_k\}$ given by the expression:
\begin{equation}
P(\{S_i,r_k\}) \sim e^{-\sum_{i,l} \beta_l R^{(l)}_i}
\sim e^{-s_{gauge}}\ ,
\label{P0}
\end{equation}
together with the statement that the integration measure is gauge
invariant too. The introduced above parameters $\beta_l$ are measures of
the uncertainty of the corresponding plaquettes. The sums run over time
moments $i$ and types of the plaquettes ($l=1$ for cash-share plaquettes
and $l=2$ for cash-derivative ones).

As it was shown in Refs\cite{hep-th/9710148}, the
assumptions which have been made are equivalent to a log-normal
model for the share price walks with the distribution function
\begin{equation}
P(S(T)|S(0)) = \frac{1}{\sigma S \sqrt{2\pi T}}
e^{-(\ln (S(T)/S(0)) - (\mu -
\frac12 \sigma^2)T)^2/(2\sigma^2 T)} \ .
\label{log}
\end{equation}
Here the volatility $\sigma=1/\sqrt{2\beta}$ and the average rate of
share return $\mu=r_b-r_1$ have been introduced.

There are two points to note here. The first one is that the log-normal
distribution was derived in an absence of matter fields. These matter
fields can significantly change the form of the distribution function and
other properties of the price random walk. Indeed, in the presence of
the money flows more complicated random processes for the price motion
emerge which resemble real data observations~\cite{ISfin1}. The second
point concerns other types of quasi-Brownian price motions which are
considered in mathematical finance. They can be also introduced in
the theory, making parameters, which we keep constant for the moment,
depending on price values.

Let us return to gauge fixing. Since the action $s_{gauge}$ is gauge
invariant it is possible to perform a gauge transformation which will not
change the dynamics but will simplify further calculation. In lattice
gauge theory~\cite{Creutz} there are several standard choices of gauge
fixing and axial gauge fixing is one of them. In the axial gauge an
element of the structural group is taken as something chosen on links in
the time direction and exchange rates along the "space" direction at some
particular time. This kind of gauge fixing is
convenient for the model in question. Actually this gauge has been used
in deriving (\ref{log}).

We choose $r_b$ to be the risk free interest rate and $r_b-r_1$,
$r_b-r_2$ are the average rates of return on the share and the
derivative. This means that in the situation of the double ladder base
the only dynamical variables are the exchange rates (prices) as a
function of time and the corresponding measure of integration is the
invariant measure $\frac{\mbox{d}S_i\mbox{d}C_i}{S_i C_i}$. Below we fix
the price of the shares at time $t=0$ taking $S_0=S(0)$. We also fix
the exchange rate of the derivative to the share at the moment of the
derivative exercise.

Let us note that quantities of our theory --- exchange and interest rates
are not gauge invariant but gauge covariant. So it is natural to
choose the gauge in which the exchange and interest rates take their real
value. In our gauge "rate of return" on cash takes its
real value $r_b$ while average rates of return on the share and the
derivative are not $r_1$ and $r_2$ but $r_b-r_1$ and $r_b-r_2$. This
provides the axial gauge fixing together with fixing the price of the shares 
at the time $t=0$ and
the exchange rate of the derivative to the share at 
the  time of exercise.

Let us return to expression (\ref{P0}). It is derived under the assumption
that arbitrage opportunities are uncorrelated between different "space-time"
plaquettes. However, there are many important problems where the correlation
between returns on elementary plaquettes operations have to be taken into
account. The most important such example is the portfolio theory where an
optimal portfolio is constructed using correlation between
assets~\cite{Blake,portfolio}. Another example is derivative pricing
which is studied in the paper.

To account for the correlation we have to elaborate some details of the 
construction of the action (the definition of the probability finding particular
configuration of exchange and interest rates). Of course we retain base
principles such as gauge invariance. The configuration which has less
arbitrage is more probable. But we look more precisely at the
problem of independence of arbitrage operations.

The general form of the action is
\begin{equation}
s_{gauge} = \sum_{\xi\zeta} Q_\xi A_{\xi\zeta} Q_\zeta /(2\Delta)
\label{gen-act}
\end{equation}
where $Q_\xi$ are local (dependent) arbitrage plaquette quantities and
matrix $A_{\xi\zeta}$ is the correlation matrix of the plaquettes.
This scheme is general and can be applied to the case of many correlated
assets. Expression (\ref{gen-act}) is gauge invariant and naturally
generalize (\ref{P0}). In general, matrix $A_{\xi\zeta}$ is not diagonal 
due to
presence of correlations .

To simplify the model we make a locality assumption. It means that
the virtual arbitrage opportunities emerging 
at different times are independent.
This makes matrix $A$ diagonal in the time index. 
The action can be rewritten in
following form:
\begin{equation}
s_{gauge} = \sum_{ill'}
Q^{(l)}_i A_{ll'} Q^{(l')}_i /(2\Delta)
\label{action-general}
\end{equation}
where $A_{ll'}$ is equal-time plaquette correlation matrix and $l$ and $l'$
run over all elementary plaquettes at fixed time. It is straightforward
to show that in the continuous limit ($\Delta\to0$) the previous expression
takes the following form
\begin{equation}
s_{gauge} = \int_0^T\sum_{ll'}
\left(\frac1S \frac{dS_l(t)}{dt} - (r_b - r_l)\right)
\frac{A_{ll'}(t)}{2}
\left(\frac1S \frac{dS_{l'}(t)}{dt} - (r_b - r_{l'})\right)dt
\ .
\label{action-general-cont}
\end{equation}
This implies the following expression for
the correlation matrix $A$:
\begin{equation}
A^{-1}_{ll'}(t) = 
<\frac1S dS_l(t), \frac1S dS_{l'}(t) >/dt \equiv \sigma^2_{ll'}(t)
\label{A}
\end{equation}
(terms $(r_b-r_l)dt$ do not contribute in continuous limit). For the
share-cash-derivative system we need to consider the following correlators:
$<\frac{1}{S(t)} dS(t), \frac{1}{S(t)} dS(t)>/dt$, 
$<\frac{1}{C(t)} dC(t), \frac{1}{S(t)} dS(t)>/dt$, 
$<\frac{1}{C(t)} dC(t), \frac{1}{C(t)} dC(t)>/dt$.
The first one is equal to $\sigma^2 \equiv 1/(2\beta_1)$.
The second correlator we denote as $\alpha(t)/(2\beta_1)$ introducing 
the notation:
\begin{equation}
\alpha(t) \equiv <\frac{1}{C(t)} dC(t), \frac{1}{S(t)} dS(t)>/
<\frac{1}{S(t)} dS(t), \frac{1}{S(t)} dS(t)>
\ .
\label{alfa}
\end{equation}
To calculate the last correlator 
$<\frac{1}{C(t)} dC(t), \frac{1}{C(t)} dC(t)>/dt $ we consider $C(t)$ 
as a random function of $S$. Then it can be represented as
$$
<\frac{1}{C(t)} dC(t), \frac{1}{C(t)} dC(t)>/dt =
\alpha^2(t)/(2\beta_1) + 1/(2\beta_2)
$$
where the term $\alpha^2(t)/(2\beta_1)$ is responsable for the 
volatility of the derivative due to its dependence 
on the share price and $1/(2\beta_2)$
stands for the volatility of 
the derivative due to the random character of function $C$ itself.
Given these parameterization with $\alpha$, $\beta_{1,2}$ we can 
find matrix $A$ from Eq(\ref{A}) and
get the following expression for the action (\ref{action-general-cont}) in this
particular derivative-related setting:
\begin{equation}
\begin{array}{rl}
s_{gauge} =&
\beta_1 \int_0^T dt
\left(\frac{1}{S(t)}\frac{dS(t)}{dt} + (r_1 - r_b) \right)^2 +
\\&+
\beta_2 \int_0^T dt
\left(
\left(\frac{1}{C(t)}\frac{dC(t)}{dt} + (r_2 - r_b) \right)
- \alpha(t)
\left(\frac{1}{S(t)}\frac{dS(t)}{dt} + (r_1 - r_b) \right)
\right)^2
\ .
\end{array}
\label{action-cont}
\end{equation}

It is interesting to note that this expression gives the main 
order (with respect to $\Delta$) of the lattice action
\begin{equation}
s_{gauge} =
\sum_i (\beta_1 R^{(1)}_i + \beta_2 R^{(2,1)}_i)
\ .
\label{action-discr}
\end{equation}
where $R^{(2,1)}$ is defined as
\begin{equation}
R^{(2,1)}_i = (Q^{(2)}_i - \alpha_i Q^{(1)}_i)^2/\Delta \ .
\label{p21}
\end{equation}
Action (\ref{action-discr}) has a very simple interpretation.
As we already mentioned, local cash-share and cash-derivative arbitrage
operations (at the same time) clearly are not independent and
we cannot use $R^{(1)}$ and $R^{(2)}$ simultaneously to characterize
the independent arbitrages. 
Instead, we can find statistically independent combinations of $Q^{(1)}_i$ and
$Q^{(2)}_i$ and define $R^{(2,1)}$
as in (\ref{p21})
to use together with $R^{(1)}_i$ as independent plaquette quantities in 
expression (\ref{P0}). This returns us to the action (\ref{action-discr}).
This is the action we deal with in the paper when we consider a lattice system
while we use action (\ref{action-cont}) for the continuous limit.

\vspace{10mm}

All said above is valid for any derivative contract. A particular type
of derivative is reflected by details of the construction of the base
graph of the theory. We consider here in more detail
European and American call
options. Other kinds of derivatives require other base constructions
which, however, are straightforward.

We start with the European call option (see Appendix 1 for the definition).
As we said before, the "space"-time graph for it and the underlying share
is the double ladder. The only new element here, comparing with the
model of Ref~\cite{ISfin1} adopted for two types of shares, is the
additional link at the expiration date between the option and cash and
the share with the direction chosen to describe the possibility to change
this option and $E$ units of cash on the share. (Fixed exchange rate on
this link gives missing gauge fixing condition.)

The graph for the American call option and the underlying share contains
the double ladder and the directed links from the option and cash to
share at any intermediate times before the expiration time $T$. These
links are the same as that for European option.

The links which are responsible for the swap of the derivative to the
share, i.e. exercising of the option, generate new plaquettes on the
"space"-time graph. Corresponding plaquette quantities $R'_i$ have to be
taken into account. These terms depend very much on concrete
conditions of the derivative contract. We keep the terms
terms aside and rewrite (\ref{P0}) in the form
\begin{equation}
P(\{S_i,r_k\}) \sim e^{-s_{gauge}-\sum_i\beta'_i R'_i}\ .
\label{P'}
\end{equation}
Below we do not concentrate on the primed plaquettes. They contain directed
links and contribute to the boundary conditions only.
However, to make complete analytical and numerical analysis
they are important and have to be retained. We return to this point 
in section 3 where we derive the boundary conditions. In the next section
we show how this formalism reproduces all the results of standard
derivative pricing theory.

\section{Derivation of the Black-Scholes equation}
In this section we obtain the Black-Scholes equation as the equation of
the saddle-point in the quasi-classical limit of the gauge theory in
absence of money flows.

Let us return to Eqn(\ref{action-cont}) for the action in the continuous
($\Delta\to0$) limit. The first term in the RHS of this equation corresponds
to geometrical random walks and provides a background for the derivation.

Being a derivative instrument from the share price, the price of the
derivative has to be correlated or even defined by the share price. That
is why it is natural and more convenient for our purposes to write $C(t)$
as a some unknown function of $S(t)$. Since we integrate over all $C(t)$
(or over all functions $C(t,S)$) this does not mean any loss of
generality. We use the fact that $C$ is a function of $S$ and the
property of the geometrical random walk of the underlying asset to
obtain a compact expression for the parameter $\alpha(t)$ (\ref{alfa}):
\begin{equation}
\alpha = \frac SC\frac{\partial C}{\partial S} \ .
\label{alpha}
\end{equation}
To explain this we use the following fact for the geometrical Brownian motion
known as Ito's lemma:
\begin{equation}
d f(t,S) = \frac{\partial f}{\partial t}dt + \frac{\partial f}{\partial S} dS
+ \frac{S^2\sigma^2}{2}\frac{\partial^2 f}{\partial S^2} dt + o(dt)
\label{Ito}
\end{equation}
where $f(t,S)$ is an arbitrary function of $t$ and $S$
(the function is supposed to be smooth)~\cite{Wilmott}.
Using this equality and the fact $<dS/S,dS/S>\to S^{-2}<dS,dS>$
we obtain Eqn(\ref{alpha}).

Now let us turn our attention to the second term in Eqn(\ref{action-cont}).
Using expression (\ref{r2}) and Ito's lemma (\ref{Ito}) we end
with the following action term:
\begin{equation}
\beta_2 \int dt
\left(\frac1C \frac{\partial C}{\partial t} +
\frac1C \frac{\sigma^2}{2} S^2 \frac{\partial^2 C}{\partial S^2} -
r_b (1 - \frac SC\frac{\partial C}{\partial S})
\right)^2 \ .
\label{aa}
\end{equation}
This term corresponds to a virtual arbitrage account. 

The classical limit
for the action, i.e. $\beta_2\to\infty$, reduces the functional
integration over functions $C(t,S)$ to {\bf the contribution from the
classical trajectory only and this trajectory is defined by the
Black-Scholes equation} for the price of the financial derivative:
\begin{equation}
\frac{\partial C}{\partial t} +
\frac{\sigma^2}{2} S^2 \frac{\partial^2 C}{\partial S^2} -
r_b (C - S\frac{\partial C}{\partial S}) = 0 \ ,
\label{BS}
\end{equation}
since, as it is easy to see, 
the rate $r_2$ on the derivative  whose
price is defined by (\ref{aa}) is give by the expression 
\begin{equation}
r_2 = r_1 <\frac SC \frac{\partial C}{\partial S}>
\ .
\label{r2}
\end{equation}

The equation does not depend on a particular type of the derivative which
is encoded in the boundary conditions. For a comparison we give standard 
no-arbitrage derivation of the
Black-Scholes equation in the
appendix.

\subsection{Boundary conditions}

In this subsection we consider boundary conditions which characterize
a particular derivative. Here we treat only European and American calls
but it is straightforward to generalize the consideration.
To this end we return to the construction of the base graph in section 2.

Let us first consider in more detail the European call option.
At the expiration time $T$ the option can be sold or bought for $C(T)$
units of cash but it can also be exchanged (together with $E$ units of
cash)
on the share.
If the price of the share $S$ is less than $E$ this leads
to effective scrap of the call.

These opportunities create new arbitrage possibilities which we take into
account by introducing the $\beta'R'_N$ term in action (\ref{P'}).
There exists a new arbitrage operation which is available 
at expiration time $T$:
one can borrow the portfolio consisted of option 
and $E$ currency
units, exchange it on the share if $S(T)>E$ 
(directed link on the base graph), sell share for $S(T)$ units
of money 
and buy the portfolio again. This gives us a plaquette excess return
$$
Q'_N = (S(T)\theta(S(T)-E) + E\theta(E-S(T)))(C(T)+E)^{-1} - 1\ ,
$$
and for the plaquette quantity $R'_N$ we get
\begin{equation}
R'_N = \frac{E\theta(E-S(T)) + S(T)\theta(S(T)-E)}{C(T)+E} +
\frac{C(T)+E}{E\theta(E-S(T)) + S(T)\theta(S(T)-E)} - 2\ .
\label{boundary}
\end{equation}
In the quasi-classical
(no-arbitrage) limit we have $\beta'\to\infty$ which
gives the following boundary condition for the
European call option from Eqn(\ref{boundary}):
\begin{equation}
C(T) = (S(T)-E)\theta(S(T)-E) \ .
\label{eur-bound}
\end{equation}
If we neglect this arbitrage possibilities at the time $T$, we can use this
equation as a gauge fixing condition and need not consider additional links
and boundary plaquettes. This approximation is quite realistic and 
clearly simplifies the scheme.

Let us turn to the American call option. It can be exchanged 
for the share at any time up to $T$ by paying additional $E$ units cash.
In this case the base graph contains links from the option to the share (but
not backwards) at all time points. It allows two possible
arbitrage operations. The first operation is similar to considering
above
operation with a European option at exercise time $T$. 
Now this operation is available at any moment $t\leq T$.
One can borrow the portfolio consisting of an option and $E$ currency units,
exchange it on the share, sell the share for $S(t)$ units of money and buy
the portfolio again. The excess return on this operation is
$$
Q'_t = S(t) (C_A(t)+E)^{-1} - 1\ ,
$$
and for the boundary plaquette quantity we get
\begin{equation}
R'_t = \frac{S(t)}{C_A(t)+E} +
\frac{C_A(t)+E}{S(t)} - 2\ .
\label{R'-am1}
\end{equation}
In close analogy with the European option in the quasi-classical (the absence
of the arbitrage) limit we get the following boundary condition for the
American call option:
\begin{equation}
C_A(t) = S(t)-E
\ .
\label{am-bound2}
\end{equation}
Now we have to determine moment when American option exercised. To this end we
consider another possible arbitrage operation.
Let an arbitrageur have portfolio of an option and
$E$ units of cash at some time $t$. He can exchange the portfolio for a share and keep it up
to time $t+dt$ or hold the portfolio and exchange it on a share at time
$t+dt$. The expected return on the portfolio is
$(C_A(t)(r_b-r_2)+Er_b)/(C_A(t)+E)$ so at time $t+dt$ an arbitrageur will
have 
$$
[1+\frac{C_A(t)(r_b-r_2)+Er_b)}{C_A(t)+E}dt]\frac{C_A(t)+E}{C_A(t+dt)+E}
$$ 
portfolios of an option and $E$ cash units
which he can exchange for the equal number of shares. The expected return on
share is $(r_b-r_1)$. Therefore the excess return on this operation is
$$
Q''_t = \biggl[1+\frac{C_A(t)(r_b-r_2)+Er_b)}{C_A(t)+E}dt\biggr]
\frac{C_A(t)+E}{C_A(t+dt)+E}
\left( [1+(r_b-r_1)dt] \frac{S(t)}{S(t+dt)} \right)^{-1} \ ,
$$
and for the boundary plaquette quantity we obtain the expression:
\begin{equation}
R''_t = Q''_t + (Q''_t)^{-1} - 2\ .
\label{R''-am1}
\end{equation}
In the no-arbitrage limit we have
$$
Q''_t -1 = 0
$$
at the exercise time. This gives us second boundary condition for the American 
call:
\begin{equation}
\frac{\partial C_A(t,S)}{\partial S} = 1
\ .
\label{am-bound1}
\end{equation}
To prove this expression we have to use the no-arbitrage condition $Q''_t -1= 0$,
expression
for $r_2$ (\ref{r2})
$$
r_2 = \frac{S}{C}\frac{\partial C}{\partial S} r_1
$$
and to take into account first boundary condition (\ref{am-bound2}).

So finally we obtain the following boundary conditions for
the American call option~\cite{Wilmott}:
\begin{equation}
\frac{\partial C_A(t,S)}{\partial S}=1
\ ,\qquad{C_A}(t,S) = S(t) - E
 \ .
\label{am-bound}
\end{equation}
The equations determine the time when the option is exchanged for the
share and the corresponding payoff. It is easy to show from the 
plaquette analysis that in absence of dividends American call is never 
exercised early and, hence, is equivalent to European call but we do not 
stop for it.

Summing up, we have shown that in the absence of money flows and in
the quasi-classical limit (i.e. suppression of the arbitrage operations with
the derivatives) the Black-Scholes equation emerges as the equation for the
saddle-point which is provided with appropriate boundary conditions.

\subsection{Connection with Black-Scholes analysis}

In conclusion we want to clarify the connection with the original
Black-Scholes analysis (see Appendix). Let us return back to the formula for
$R^{(2,1)}$ (\ref{p21}):
$$
R^{(2,1)}_i = (Q^{(2,1)}_i)^2/\Delta
$$ 
with 
$Q^{(2,1)}_i$ defined as
$$
Q^{(2,1)}_i\equiv Q^{(2)}_i - Q^{(1)}_i \alpha_i \  .
$$ 
To first order in $\Delta$ the last expression can be rewritten as
$$
Q^{(2,1)}_i =
C_i S_i^{-\alpha_i} e^{\alpha_i r_1\Delta} e^{(1-\alpha_i)r_b\Delta}
e^{-r_2\Delta} S_{i+1}^{\alpha_i} C_{i+1}^{-1} - 1 \ .
$$
It is easy to give a simple interpretation of the last expression.
Indeed, it is not difficult to see that it describes the following 
money circle:
$$
C \to USD \to (1-\alpha) USD \oplus \alpha S
\to USD \to C \ .
$$
The portfolio comprising shares and cash (in parts $\alpha$ and $1-\alpha$)
emerging at the intermediate state is the Black-Scholes hedging portfolio 
for the derivative and
the condition (\ref{correl}) is exactly 
the hedging relation. From this point of
view the plaquette $Q^{(2,1)}_i$ represents the arbitrage
fluctuations in hedging portfolio-derivative plaquette. This
returns us back to the treatment of the expression 
$$
(\frac{\partial
C}{\partial t} + \frac{\sigma^2}{2} S^2 \frac{\partial^2 C}{\partial^2 S}
- r_b (C - S\frac{\partial C}{\partial S}))/ 
(C- S\frac{\partial C}{\partial S})
$$
as the arbitrage excess return on the infinitesimal operation~\cite{Wilmott}.

\section{Money flows and correction to the Black-Scholes equation}

Now let us turn to money flow fields. These fields represent cash-debt
flows in the market. The importance of the cash-debt flows for our
consideration is explained by their role in a stabilization of market
prices. Indeed if, say, some asset price creates a possibility to get
a bigger return than from other assets (with similar risk), 
then an effective cash flow appears, directed to these more valuable 
shares. This causes restoration of equilibrium
due to the demand-supply mechanism. The same picture is valid for debt flows
if there is a possibility for debts restructurisation. As we will see all
these features find their place in the GTA framework.

Following Refs~\cite{hep-th/9710148} we can formulate the dynamics
for the cash-debts flows basing on several assumptions, such as gauge
invariance of the dynamics, an investor's wish to maximize his return and
his limited rationality. This gives us the following functional integral
representation for the matrix element of an evolution operator in the
coherent state representation of the money flows in the case of our
double ladder base graph ($r_0\equiv r_b$):
\begin{equation}
<\bar{\psi}_N,\bar{\chi}_N|
\hat{U}(t+N\Delta,t)|\psi_0,\chi_0> =
\int \prod_{k=0,i=1}^{2,N-1}d\bar{\psi}_{k,i} d\psi_{k,i}
d\bar{\chi}_{k,i}d\chi_{k,i}
e^{(s1 + s1'+ s_b)} \ ,
\label{int}
\end{equation}
with the actions for cash and debt flows:
$$
s1= \sum_{k=0,i=0}^{2,N-1} (
\bar{\psi}_{k,i+1} e^{\beta r_k\Delta} \psi_{k,i} -
\bar{\psi}_{k,i}\psi_{k,i})
+ \sum_{i=0}^{N-1} (
(1-t_c)^{\beta}C_i^{\beta} \bar{\psi}_{0,i+1}\psi_{2,i}
$$
\begin{equation}
+ (1-t_c)^{\beta} C_i^{-\beta} \bar{\psi}_{2,i+1} \psi_{0,i})
+ (1-t_c)^{\beta}S_i^{\beta} \bar{\psi}_{0,i+1} \psi_{1,i}
+ (1-t_c)^{\beta} S_i^{-\beta} \bar{\psi}_{1,i+1} \psi_{0,i}) \ ,
\label{s1}
\end{equation}
$$
s1' = \sum_{k=0,i=0}^{2,N-1} (\bar{\chi}_{k,i+1}
e^{-\beta r_k \Delta} \chi_{k,i}
- \bar{\chi}_{k,i}\chi_{k,i}) +
\sum_{i=0}^{N-1} ((1+t_c)^{-\beta}C_i^{-\beta} \bar{\chi}_{0,i+1} \chi_{2,i}
$$
\begin{equation}
+ (1+t_c)^{-\beta}C_i^{\beta} \bar{\chi}_{2,i+1} \chi_{0,i}
+(1+t_c)^{-\beta}S_i^{-\beta} \bar{\chi}_{0,i+1} \chi_{1,i}
+ (1+t_c)^{-\beta}S_i^{\beta} \bar{\chi}_{1,i+1} \chi_{0,i} ) \ .
\label{s1'}
\end{equation}
Here $t_c$ is a relative transaction cost and the action $s_b$ represents
exchanges along boundary plaquettes and is contract-dependent.
Using this expression it is easy to obtain the transition probability
in the occupation number representation simply by integrating over
$\bar{\psi}$, $\psi$, $\bar{\chi}$, $\chi$
variables.
Let us consider a transition to the state
with $n_0 (n'_0)$ cash units in
the
the long (short) position, $n_1 (n'_1)$ stock units in
the long (short) position, $n_2 (n'_2)$ derivatives units in
the long (short) position at time $t_1=t+N\Delta$
from the state ($m_k$,$m'_k$) at the original time $t$.
The long position corresponds to possessing assets and
the short position implies possessing liabilities ("debts").
The probability for this transition
has been derived in~\cite{ISfin1}:
$$
P((\{n_k\},\{m_k\}),(\{n'_k\},\{m'_k\}),t_1,t) =
S^{\beta (m_1 -m'_1 -n_1+n'_1)}(t)
C^{\beta (m_2 -m'_2 -n_2+n'_2)}(t)
$$
\begin{equation}
\int d\psi d\bar{\psi} d\chi d\bar{\chi}
<\bar{\psi}_N,\bar{\chi}_N|
\hat{U}(t+N\Delta,t)|\psi_0,\chi_0>
e^{-\bar{\psi}_{N}{\psi}_{N}}
\prod_{k=0}^{2} \bar{\psi}^{m_k}_{k,0} \bar{\chi}^{m'_k}_{k,0}
{\psi}^{n_k}_{k,N} {\chi}^{n'_k}_{k,N} /{n_k!n'_k!} \ .
\label{int1}
\end{equation}
Expressions (\ref{action-discr},\ref{P'},\ref{int},\ref{int1}) form the
complete set of necessary equations to describe the dynamics and mutual
influence of interest/price rates and money fields. 

We now derive the correction to the
effective action of the gauge field due to the presence of money flows.
Since, as it was shown in the previous section, the Black-Scholes
equation appears in the quasi-classical limit of the free gauge theory,
the correction to the effective action leads in the same limit to a
correction to the Black-Scholes equation.

Let us suppose that the initial asset configuration at time $t$
is given by the probability distribution $F(\{m_k\},\{m^{\prime}_k\})$.
Then, as it follows from section 2, the correction to the
effective action for the gauge field is equal to:
$$
\delta s_g = \ln\sum_{\{n_k\},\{n'_k\},\{m_k\},\{m'_k\}}
P((\{n_k\},\{m_k\}),(\{n'_k\},\{m'_k\}),0,T)
F(\{m_k\},\{m^{\prime}_k\})
$$
where the function
$P((\{n_k\},\{m_k\}),(\{n'_k\},\{m'_k\}),0,T)$
is given by Eqns(\ref{int} - \ref{int1}).
The generalization of this expression
to many investment horizons is straightforward but cumbersome.

The correction to the effective action of the prices' gauge field
transforms the Black-Scholes equation for the classical trajectory to the
following equation for the trajectory $C(t)$ minimizing the action:
\begin{equation}
min_{C(\cdot)} \left(
\int dt
\left(\frac1C \frac{\partial C}{\partial t} +
\frac1C \frac{\sigma^2}{2} S^2 \frac{\partial^2 C}{\partial^2 S} -
r_b (1 - \frac SC\frac{\partial C}{\partial S})
\right)^2 - \frac{1}{\beta_2}\delta s_g \right) \ .
\label{BS1}
\end{equation}
Then the Black-Scholes equation is formally substituted by the equation:
\begin{equation}
\frac{\delta}{\delta C(t)} \left(
\int dt
\left(\frac1C \frac{\partial C}{\partial t} +
\frac1C \frac{\sigma^2}{2} S^2 \frac{\partial^2 C}{\partial^2 S} -
r_b (1 - \frac SC\frac{\partial C}{\partial S})
\right)^2 - \frac{1}{\beta_2}\delta s_g \right) = 0
\label{BS2}
\end{equation}
with the same boundary conditions. However, we think that the former form
of the problem (Eqn(\ref{BS1})) is more convenient for numerical solution
while the latter (Eqn(\ref{BS2})) cannot provide much for analytical
inside since even in the more simple situation of the GTA model of stock
exchange~\cite{ISfin1} an analytical approach hardly produce valuable
results.

However we can state general properties of the correction which can be
derived in the same way as was done for the correction to the
effective action for the price of shares~\cite{ISfin1}. It was shown
there that the correction from money flows has two important properties:
\begin{enumerate}
\item The correction vanishes at large time intervals;
\item The correction disappears
in a limit of completely noisy traders who do not consider their
potential profit as a motivation for transactions, i.e.
in the limit $\beta\to0$;
\item The correction is governed
by a number of traders (read as money available for arbitrage operations),
i.e. the correction disappears when no money is available for arbitrage
operations.
\end{enumerate}
This results in the convergence of the price movement process to the
geometrical Brownian walks at large time scale which is well-know from
real data observation. "Physically" it means that for large times
the arbitrage
is washed out (if there are arbitrageurs and
they prefer to get more than less), no-arbitrage
constraint holds firm and the Black-Scholes
description is correct.

We state the central result of the paper as a derivation 
of the Black-Scholes equation and the corresponding corrections.
However we want to emphasize that {\it to
account for both the virtual arbitrage and the money flows one needs to
calculate the functional integral over money fields and prices keeping
all action terms, in particular boundary plaquettes}.

\section{Conclusion}
In conclusion, in the paper we have shown how to adapt the Gauge Theory of
Arbitrage to apply it to derivative pricing. This can be done by constructing
base
graphs for the theory which allows exchange of the derivatives and the
underlying shares not only for cash but also for each other at some
prespecified moments of time. The framework is general enough to describe
both European and American derivatives with various payoffs. It was
demonstrated that in a quasi-classical limit in absence of money flows,
the treatment reproduces the Black-Scholes equation with appropriate
boundary conditions. Formal expression for the corrections to the equation 
in the presence of money flows is also obtained.

All listed above results are obtained analytically. However, any further
developments give rise to some quite complicated calculations. We believe
that only computer calculation can demonstrate the agreement (or
disagreement) of the theory with financial data and further development
of the proposed approach to the financial modelling should be based on
computer simulation. Indeed, as we mentioned before, even in the much
easier case of cash-and-shares system only results of numerical
calculation can be successively compared with real statistics. The
situation becomes more extreme with the derivative pricing. On the other
hand, the numerical algorithms to be used in this context essentially the
same as ones, which have been used in lattice gauge theory with matter
fields.

\section*{Acknowledgments}
We want to thank Alexandra Ilinskaia for many helpful discussions.
We are grateful to Paul Cooper for careful reading of the manuscript and
to Jeff Miller who let us know about Ref~\cite{Hughston}.
This work was supported by IPhys Group and grant UK EPSRC GR/L29156.

\section*{Appendix 1}
In this appendix we give a simple introduction to the financial derivatives
and an almost rigorous derivation of the Black-Scholes equation using
standard no-arbitrage arguments.

A financial derivative is a financial instrument whose value is defined
by values of other (underlying) financial variables. These underlying 
variables may be prices of
stocks, bonds or other derivatives, exchange rates, market indices and so
on. Some particular derivatives (as warrants) have been known for
centuries, but after the introduction in 1973 of exchange-traded
derivatives on stocks  in US trading in derivatives became a really huge
industry (for practical aspects see, for instance,~\cite{Dubofsky,Hull}).
Most popular derivatives are call and put options, futures while swaps,
forwards and other derivatives are also important for practitioners. Let
us give a simple  definition of these derivatives.

The call option gives a right but not an obligation to buy a certain
number of shares (or any prespecified underlying asset) at fixed (strike)
price $E$. If the right can be used at the {\it final} moment  of
(expiration) time $T$ only, the call option is called European. In
contrast, the American call option provides the right to buy the
underlying asset at the strike price at {\it any} time before the
expiration time $T$. The put options guarantee the same rights not for
buying but for selling of  the share. Since the American option gives an
investor more freedom it is  more valuable. In general, the price of the
options are nonzero and depend on the price of the underlying asset. The
only time when the option price can be zero is the expiration time
subject to the case when the underlying asset price is less than 
$E$ (for call options)
or more than $E$ (for put options).

A futures contract is an agreement between two parties to buy or sell an
asset at a certain time in the future for a certain price. This agreement
is  a security and can be traded. The price of the agreement clearly
depends on  the underlying asset price and can be positive as well as
negative. The  negative price means that to close the position an
additional amount should be paid. There is a number of methods to
estimate derivative prices and hedge them (which is even more important
from practical point of view). The most  standard is Black-Scholes
analysis which is based on no-arbitrage condition and quasi-Brownian
character of the underlying price. Below we assume that the underlying
asset is a share and reproduce simple version using the geometrical
Brownian process for the underlying share price:
$$
dS/S = \mu dt + \sigma dW
$$
where $dW$ is the Wiener process, $\mu$ is an average rate of return
on the share and $\sigma$ is a standard deviation of the return (so
called volatility). Though our derivation is not strictly speaking
rigorous, a formal derivation can be found in~\cite{Duffie}. The
derivative price $C$ is determined by the share price $S$. The share
price  at some moment in the future depends only on the current share price,
not  on the share price in the past (i.e. we assume a form of
market efficiency). We also assume that the derivative price $C$ is not
influenced by other factors. This means that $C(t,S)$ is nonstochastic
function of the stochastic parameter $S$ which means that for given $t$
and $S$, $C(t,S)$ has a definite value. Using Ito's lemma  (\ref{Ito})
for the geometrical Brownian process, we get following equation
describing the derivative price movement
\begin{equation}
d C(t,S) =
\left(\frac{\partial C}{\partial t} +
\frac{S^2\sigma^2}{2}\frac{\partial^2 C}{\partial S^2} \right) dt +
\frac{\partial C}{\partial S} dS + o(dt) \ . \label{option-move}
\end{equation}
RHS of the equation contains $dS$ so option price movement
is stochastic. However, we can construct a portfolio from the derivative
and the underlying shares which will be risk-free. Indeed, let us
consider portfolio of the derivative and ${\partial C}/{\partial S}$
shares in the short position. Price movement of the  portfolio $\Pi$ is
\begin{equation} d\Pi =
\left(\frac{\partial C}{\partial t} +
\frac{S^2\sigma^2}{2}\frac{\partial^2 C}{\partial S^2} \right) dt + o(dt)
\ .
\label{portfolio-move}
\end{equation}
Here we omit the term $Sd[{\partial C}/{\partial S}]$ which describes a
change of the portfolio structure but not the portfolio value. Since in
the last equation there is no longer $dS$ term, the portfolio is risk-free
and cannot grow faster than the risk-free interest rate on a bank deposit.
This latter statement is known as the no-arbitrage condition 
(the violation of the
condition we consider in the main text). 
This gives us the following
equation:
$$
\frac{\partial C}{\partial t} +
\frac{S^2\sigma^2}{2}\frac{\partial^2 C}{\partial^2 S} =
\left(C-S\frac{\partial C}{\partial S}\right) r_b
$$
which is the famous Black-Scholes equation. It describes any
derivatives and the specific of a concrete derivative is encoded in
boundary conditions only. For example, the boundary condition for the
futures contract is
$$ C(S,T) = S-E $$
while for the European call it takes the form:
$$ C(S,T) = (S-E) \theta (S-E) \ .$$
Other types of boundary conditions and various modifications of the
Black-Scholes equations can be found in~\cite{Hull}.

The last note here concerns transaction costs. As we consider continuous time
limit we neglect $o(dt)$ in formula (\ref{option-move}). This means that
the portfolio of share and cash (hedging portfolio) have to be rearranged
continuously. This leads to infinite transaction costs. In this case we
have to keep finite time steps and can not neglect $o(dt)$. This results
in impossibility of perfect hedging portfolio construction.


\end{document}